# Gate-controlled field emission current from MoS$_2$ nanosheets

*Aniello Pelella, Alessandro Grillo, Francesca Urban, Filippo Giubileo, Maurizio Passacantando, Erik Pollmann, Stephan Sleziona, Marika Schleberger, and Antonio Di Bartolomeo\**

Aniello Pelella, Alessandro Grillo, Francesca Urban, Prof. Antonio Di Bartolomeo
Physics Department "E.R. Caianiello", University of Salerno, via Giovanni Paolo II 132, Fisciano 84084, Salerno, Italy
E-mail: adibartolomeo@unisa.it

Dr. Filippo Giubileo
CNR-SPIN Salerno, via Giovanni Paolo II 132, 84084, Fisciano, Italy

Maurizio Passacantando
Department of Physical and Chemical Science, University of L'Aquila, and CNR-SPIN l'Aquila, via Vetoio, Coppito 67100, L'Aquila, Italy

Erik Pollmann, Stephan Sleziona, Prof. Marika Schleberger
Fakultät für Physik and CENIDE, Universität Duisburg-Essen, Lotharstrasse 1, D-47057, Duisburg, Germany

**Abstract**

Monolayer molybdenum disulfide (MoS$_2$) nanosheets, obtained via chemical vapor deposition onto SiO$_2$/Si substrates, are exploited to fabricate field-effect transistors with n-type conduction, high on/off ratio, steep subthreshold slope and good mobility. The transistor channel conductance increases with the reducing air pressure due to oxygen and water desorption. Local field emission measurements from the edges of the MoS$_2$ nanosheets are performed in high vacuum using a tip-shaped anode. It is demonstrated that the voltage applied to the Si substrate back-gate modulates the field emission current. Such a finding, that we attribute to gate-bias lowering of the MoS$_2$ electron affinity, enables a new field-effect transistor based on field emission.

## 1. Introduction

Transition metal dichalcogenides (TMDs) have attracted a lot of attention in the past decades due to their several promising properties for electronic and optoelectronic applications. Such properties include tunable bandgap by the number of layers in the range of 0.7–2.2 eV, mobility comparable or superior to that of Si channels in modern devices, pristine interfaces without out-of-plane dangling bonds, flexibility, transparency, photoluminescence, broadband light response, thermal stability in air and high scalability for device fabrication.[1–4] TMDs are produced by mechanical or liquid exfoliation, chemical vapor deposition (CVD), molecular beam epitaxy, pulsed laser deposition, etc.[5,6]

Molybdenum disulfide (MoS$_2$) has been one of the most studied TMDs, due to its semiconducting nature and bandgap from 1.2 eV to 1.9 eV covering a range suitable for several uses.[7–13] The direct-bandgap monolayer is formed by a molybdenum (Mo) atomic plane sandwiched between two sulphur (S) planes and is bonded to other monolayers by weak van der Waals forces to form the bulk material.

Owing to the large bandgap, monolayer MoS$_2$ is a promising channel material for field-effect transistors (FETs) with high on/off ratio and fast switching. MoS$_2$ FETs show carrier mobility in the range of few dozens cm$^2$V$^{-1}$s$^{-1}$ at room temperature, which increases to few hundreds cm$^2$V$^{-1}$s$^{-1}$ when SiO$_2$ is replaced by a high-k dielectric with enhanced screening effects.[14–16] Moreover, the direct bandgap enables high optical absorption and efficient carrier excitation suitable for optoelectronic applications.[17]

The electrical properties of two-dimensional (2D) materials are sensitive to ambient gases and pressure variation and can be utilized as gas sensors.[18] It has been reported that the adsorbed gases on the MoS$_2$ channel of FETs result in degradation of device conductance, in enhanced hysteresis and in threshold voltage shifting.[14,18,19] Conversely, vacuum annealing can increase the MoS$_2$ device conductance by desorbing the gas molecules.

Due to the suitable mechanical, thermal and electronic properties, several studies have considered 2D materials in field emission devices.[20–25] TMDs possess atomically sharp edges and localized defects that can enhance the local electric field and enable the extraction of a field emission current with low turn-on voltage. Field emission (FE) is a quantum mechanical phenomenon in which electrons, extracted from a conductor or a semiconductor surface under application of an intense electric field, move in vacuum from a cathode to an anode. FE is used in a variety of applications, ranging from electrically-operated floating-gate memory cells,[26,27] electron microscopy[28] and e-beam lithography[29] to display technology[30] or vacuum electronics.[31]

Fowler and Nordheim developed a field emission theory for planar electrodes that is commonly applied also to rough surfaces where tip-shaped protrusions enhance the local electric field and emit electrons at a reduced anode-to-cathode voltage.[32] Therefore, nanostructures (nanoparticles, nanowires, nanotubes and in general 2D materials), for their high aspect ratio and intrinsically sharp edges, are ideal field emission sources.[22,33–43]

FE current from MoS$_2$ flakes with a low turn-on field and a high field enhancement factor has been reported from both the edges and the flat part of few-layer MoS$_2$ flakes.[44–47] The intrinsic n-doping of MoS$_2$, possibly controlled by a back gate,[48] has been pointed out as an important feature for future MoS$_2$ FE applications.

In this paper, we use chemical vapor deposition (CVD) to fabricate monolayer MoS$_2$ flakes on a SiO$_2$/Si substrate and characterize their transport properties. Furthermore, using a tip anode that can be accurately positioned near the edge of the flake, we investigate the local field emission properties of MoS$_2$ nanosheets. We finally demonstrate that the emission current is modulated by the gate voltage applied to the Si substrate. Our finding provides the first experimental evidence of a MoS$_2$ FET based on field emission.

## 2. Experimental section

A three-zone split tube furnace, purged with 500 Ncm³/min of Ar gas for 15 min to minimize the O$_2$ content, was used to grow MoS$_2$ flakes by CVD. The p-Si substrate capped by 285 nm thick SiO$_2$ was initially spin coated with 1% sodium cholate solution. The substrate and the MoO$_3$ precursor, obtained from a saturated ammonium heptamolybdate (AHM) solution annealed at 300 °C under ambient conditions, were placed in one of the three zones of the tube furnace, while 50 mg of S powder were positioned upstream in a separate heating zone. The zones containing the S and MoO$_3$ were heated to 150 °C and 750 °C, respectively. The growth process was stopped after 15 min and the sample was cooled down rapidly.[49]

Larger MoS$_2$ flakes were selected to fabricate field-effect transistors (FET) through standard photolithography and lift-off processes. **Figure 1(a)** and **1(b)** show the scanning electron microscope (SEM) top view of a typical device and its schematic layout and biasing circuit. The Si substrate is the back-gate while the evaporated Ti/Au (10/40 nm) electrodes are the source and the drain of the FET. The flakes are monolayer according to the photoluminescence (PL) and Raman characterization performed before metal deposition. The spectra of **Figure 1(c)** and **1(d)** show a PL peak at 1.85 eV and a separation about 20 cm$^{-1}$ between $E_{2g}^1$ and the $A_{1g}$ Raman mode typical of the monolayer. Raman measurements were performed with a Renishaw InVia microscope (Renishaw plc, New Mills, Wotton-under-Edge, Gloucestershire GL12 8JR, UK) with 532 nm excitation laser wavelength and power density below 0.1 mW/µm² to prevent channel damage.

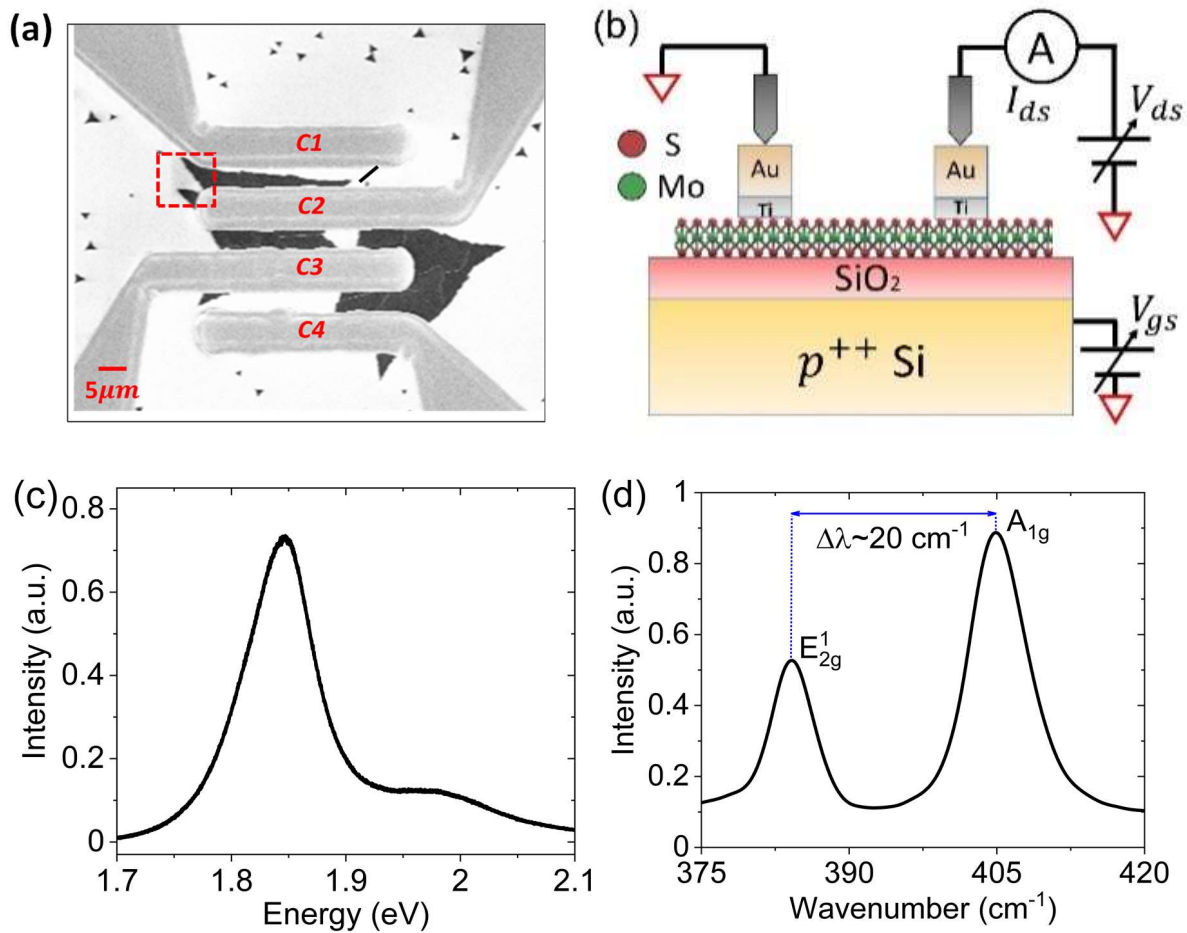

**Figure 1.** (a) SEM images of the MoS$_2$ device. (b) MoS$_2$ FET layout and schematic of the common source configuration used for the electrical characterization. (c) Photoluminescence and (d) Raman spectra of monolayer MoS$_2$ before metal deposition.

The electrical measurements were carried out inside an SEM chamber (LEO 1530, Zeiss, Oberkochen, Germany), endowed with two metallic probes (tungsten tips) connected to a Keithley 4200 SCS (source measurement units, Tektronix Inc., Beaverton, OR). Piezoelectric motors control the motion of the tips that can be positioned with nanometer precision. The chamber was usually kept at room temperature and pressure below $10^{-6}$ Torr.

In the following, the electrical characterization refers to the transistor between the contacts labelled as C2 and C3 in Figure 1(a), where two flakes in parallel are involved. The contact C3 constitutes the drain and C2 the grounded source (Figure 1(b)). The red-dashed square in Figure 1(a) highlights the edge of the $MoS_2$ flake that was used for field emission measurements with the electrode C2 as the cathode and one of the SEM tips as the anode (see the following).

## 3. Results

In high vacuum, the C2-C3 device exhibits asymmetric output characteristic, as shown in **Figure 2(a)**. A higher $I_{ds}$ drain current is measured at negative $V_{ds}$ drain voltage suggesting the formation of Schottky barriers at the contacts. The Schottky barriers are typical of $MoS_2$ devices and are caused by Fermi level pinning above the midgap of $MoS_2$.[50–52] The $I_{ds}$-$V_{gs}$ transfer characteristic of **Figure 2(b)** reveals a n-type behavior, with on/off ratio higher than five orders of magnitude, subthreshold slope SS = 7.6 V/decade and carrier mobility up to 20 $cm^2V^{-1}s^{-1}$. The n-type conduction has been attributed to the Fermi level pinning as well as to the intrinsic n-doping of $MoS_2$ due to defects such as S vacancies.[51,53,54]

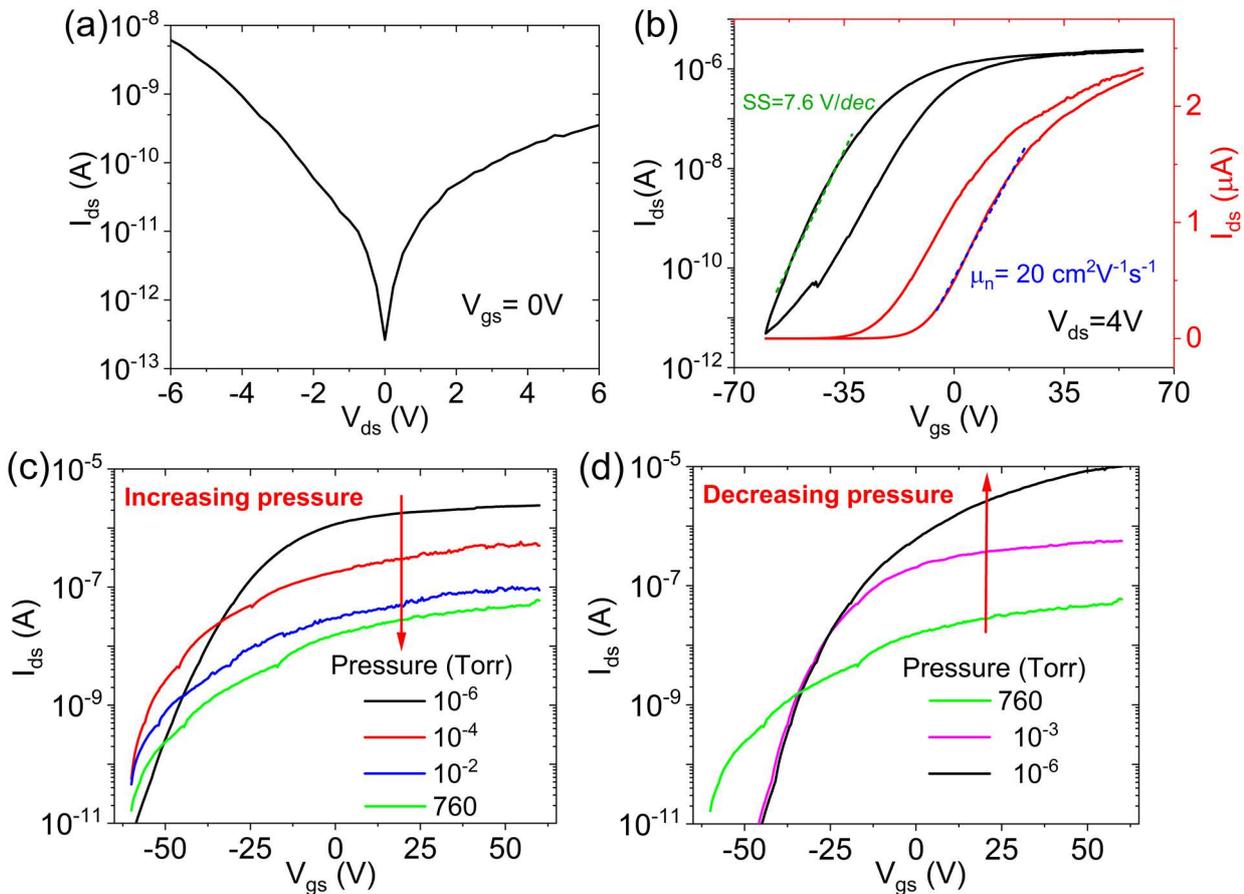

**Figure 2.** Output (a) and transfer (b) characteristics of the device between C3 and C2 contacts, with C3 used as the drain and C2 as the grounded source. The transfer characteristic is shown on both logarithmic and linear scale. Transfer characteristics for (c) increasing and (d) decreasing air pressure.

We investigated the effect of air pressure on the current flowing in the channel. **Figure 2(c)** shows a decrease of the on-current when the pressure is raised from $10^{-6}$ Torr to the atmospheric one. The effect of pressure is reversible as demonstrated by the set of curves of **Figure 2(d)**, measured for decreasing pressure. The effect of air pressure on the channel conductance has been previously investigated in transistors with $MoS_2$ as well as with other TMD materials like $WSe_2$ or $PdSe_2$.[55–57] The degradation of conductance with increasing pressure, in some cases resulting in the dramatic transformation from n-type to p-type conduction, has been mainly attributed to adsorbed oxygen and water that act as interface traps and scattering centers with a density in the order of $10^{12}$ cm$^{-2}$ eV$^{-1}$ and reduce the n-type doping of the channel.[56–60]

The intrinsic and gate-controllable n-type doping, the low electron affinity (4.2 eV),[61] and the nanosheets' sharp edges make 2D $MoS_2$ appealing for FE applications.[62,63] To perform field emission measurements, we used a tip as the anode and contact C2 as the cathode (**Figure 3(a)**). The voltage of the tip, positioned at fixed distances from the edge of the flake, was increased while monitoring the current.

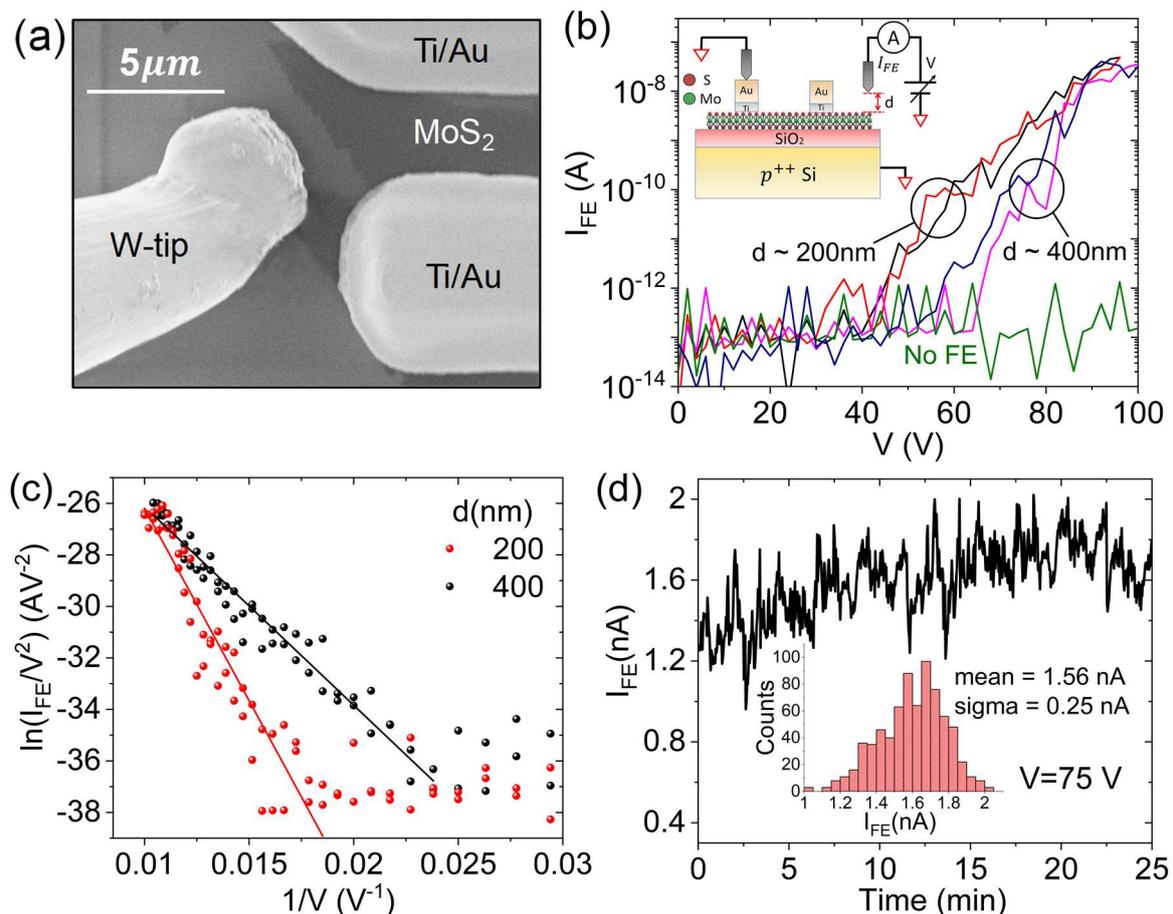

**Figure 3.** (a) SEM image of the $MoS_2$ flake (the same as in the red dashed square of Figure 1(a)) used for the measurement of field emission. The W-tip works as the anode while the Ti/Au lead is the cathode. (b) Field emission measurements performed at two different distances using the setup in the inset. (c) Fowler-Nordheim plot reveals a field enhancement factor of about 16.5 and 17.0 at distances 200 nm and 400 nm distances, respectively. (d) Field emission current stability with the tip anode at distance d=200 nm and with 75 V bias. In the inset the histogram with standard deviation of about 15%.

**Figure 3(b)** shows two sets of field emission current measurements performed at different anode-cathode distances, using the setup schematized in the inset. They show repeatable FE current occurring with a 100-120 $V\mu m^{-1}$ turn-on field (defined as the field to which the current emerges from the noise floor) and achieves a maximum value of about 20 nA.

According to the Fowler–Nordheim (FN) model, the FE current from a semiconductor can be described as:[32]

$$I_{FE} = Sa\frac{E_S^2}{\chi}e^{-b\frac{\chi^{3/2}}{E_S}} \qquad (1)$$

where S is the emitting surface area, $a = 1.54 \times 10^{-6}\ AV^{-2}$ eV and $b = 6.83 \times 10^7\ V\ cm^{-1}\ eV^{-3/2}$ are constants, $E_S\ (Vcm^{-1})$ is the electric field at the emitting surface and χ is the electron affinity of the emitting material. The electric field $E_S = \beta\frac{V}{k\cdot d}$, with β the so-called field enhancement factor, i.e. the ratio between the electric field at the sample surface and the applied field $V/(k\cdot d)$, and k~1.6 a phenomenological factor accounting for the spherical shape of the tip.[64,65] The Fowler–Nordheim equation leads to the linear behaviour of the so-called FN plot of $ln(I_{FE}\ V^{-2})\ vs.V^{-1}$ of **Figure 3(c)**, which allows estimating the β factors as 16.5 and 17.0 for 200 nm and 400 nm distances, respectively. FE current stability is demonstrated for about 25 minutes in **Figure 3(d)**, with a standard deviation of about 15%.

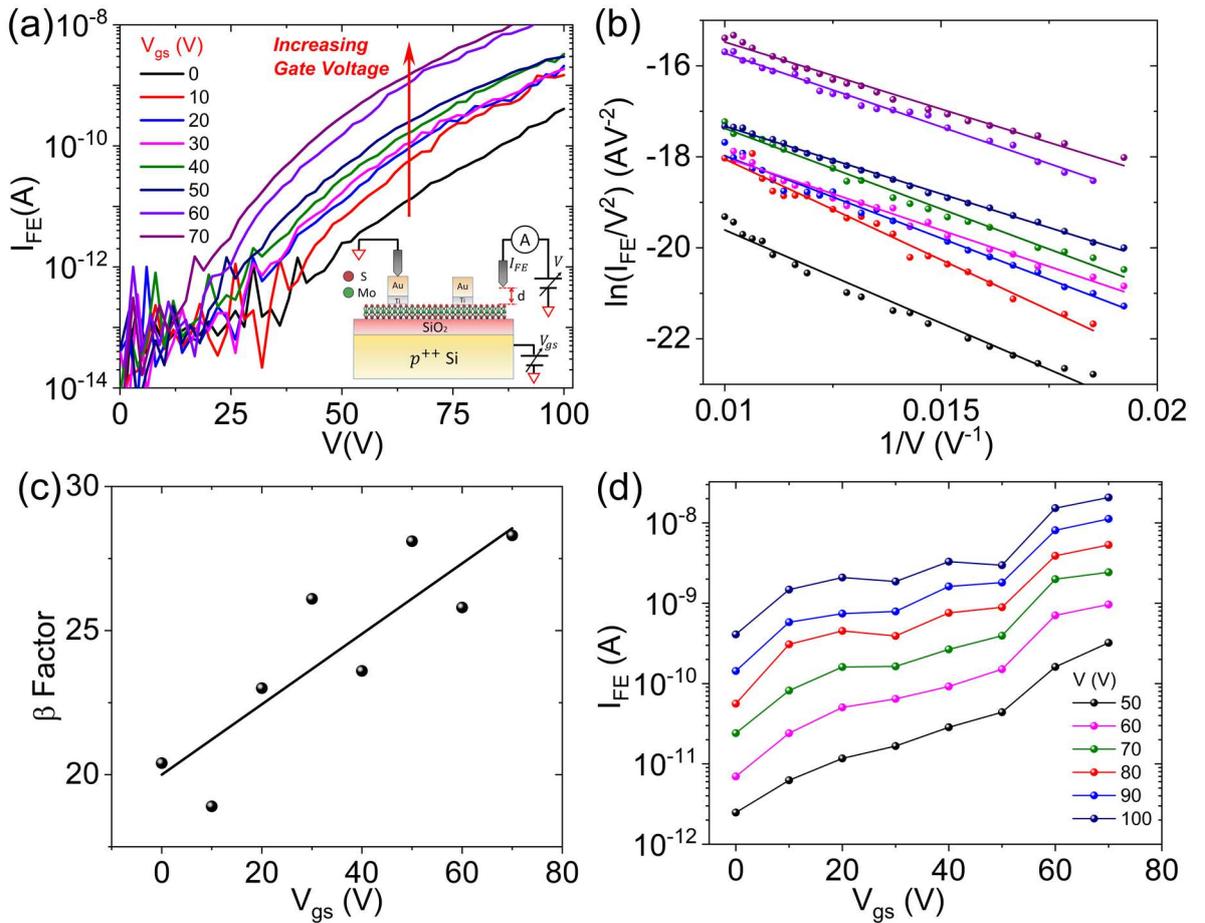

**Figure 4.** (a) Field emission current measured at d=200 nm for increasing gate voltage. A schematic of the used setup is shown in the inset. (b) Corresponding Fowler-Nordheim plot showing linear behavior of the $ln(I_{FE}V^{-2})\ vs.V^{-1}$ curves. (c) Field enhancement factor as a function of the gate voltage. (d) Transfer

characteristics at various anodic voltages V showing the exponential growth of the FE current for increasing $V_{gs}$.

The back gate can be used to electrically control the doping level of the MoS$_2$ channel. Greater availability of conduction electrons increases the tunneling probability. Therefore, a positive voltage on the gate is expected to enhance the field emission current. Indeed, **Figure 4(a)** confirms an increasing field emission current for increasing gate voltages. **Figure 4(b)** and **4(c)** show the corresponding FN plot and the extracted field enhancement factor. The growing field enhancement factor with $V_{gs}$ is an artifact related to the enhanced MoS$_2$ doping level rather than to a real field enhancement.

The data of Figures 4(a) and 4(b) provide the proof-of-concept of a new MoS$_2$ field-effect transistor based on field emission. The transfer characteristic of the device, at different anode voltages V, are shown in **Figure 4(d)**. The exponential growth of the field emission current for increasing $V_{gs}$ can be explained as follows. The applied gate and anode voltages lower the MoS$_2$ electron affinity, which can be written as:

$$\chi' = \chi - CV_{gs} \qquad (2)$$

where $\chi$=4.2 eV is the electron affinity at zero bias, and C is a constant which accounts for the gate efficiency. The band diagrams of **Figure 5(a)** schematically show the lowering of the electron affinity. Using Eq. (1) and Eq. (2), the FE current can be written as:

$$I_{FN} = A\left(\frac{1}{\chi - CV_{gs}}\right) e^{-B(\chi - CV_{gs})^{\frac{3}{2}}} \qquad (3)$$

where we have introduced the new constants $A = S a \beta^2 / (kd)^2$ and $B = bkd/\beta$ for simplicity. Eq. (3) fits the experimental data as shown in **Figure 5(b)**, confirming the consistency of the model.

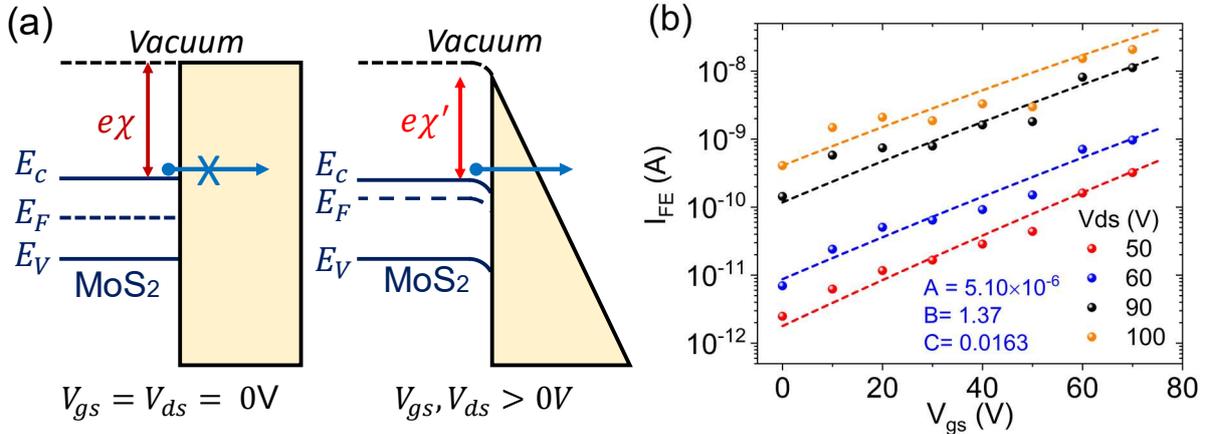

**Figure 5.** (a) MoS$_2$-vacuum band diagram at zero bias and at positive gate and anode bias. The applied voltages reduce the MoS$_2$ electron affinity as well as the barrier width (colored shapes) enabling electron tunneling. (d) Subset of data of Figure 4(d) with fit provided by eq. (3), demonstrating the consistency of the proposed model.

**Conclusions**

In summary, we have fabricated and studied monolayer MoS$_2$ back-gate field-effect transistors with n-type conductance, high on/off ratio, steep subthreshold slope and good mobility. It has been shown that the channel current is reversibly controlled by air pressure, with increase in on-current at high vacuum due to oxygen and water desorption. A field emission current, following the FN model, has been measured from the edge of the

MoS$_2$ nanosheets. More importantly, it has been shown that the gate voltage can modulate the FE current thus featuring a new transistor based on field emission. This finding constitutes a first step toward a device with great application potential, especially if implemented with the current flowing parallel to the substrate surface.

**Acknowledgements**

A.D.B. acknowledges the financial support from MIUR – Italian Ministry of Education, University and Research (projects Pico & Pro ARS01_01061 and RINASCIMENTO ARS01_01088). M.S. acknowledges the financial support from DFG - German Research Foundation (project number 406129719). The authors thank Lukas Madauß from the Universität Duisburg-Essen for contributing to sample preparation and characterization. The Authors thank ICAN - facility founded by the German Research Foundation (DFG, reference RI_00313) - for Raman and PL spectroscopy and AG Lorke for providing access to their clean room facilities.